# MatMat: Matrix Factorization by Matrix Fitting

Hao Wang
Ratidar.com
Beijing, China
haow85@live.com

*Abstract*—Matrix factorization is a widely adopted recommender system technique that fits scalar rating values by dot products of user feature vectors and item feature vectors. However, the formulation of matrix factorization as a scalar fitting problem is not friendly to side information incorporation or multi-task learning. In this paper, we replace the scalar values of the user rating matrix by matrices, and fit the matrix values by matrix products of user feature matrix and item feature matrix. Our framework is friendly to multitask learning and side information incorporation. We use popularity data as side information in our paper in particular to enhance the performance of matrix factorization techniques. In the experiment section, we prove the competence of our method compared with other approaches using both accuracy and fairness metrics. Our framework is an ideal substitute for tensor factorization in context-aware recommendation and many other scenarios.

## I. Introduction

Matrix factorization is a well-know computational approach in the field of recommender systems. The classic form of matrix factorization is a scalar fitting and dimensionality reduction problem, namely, the scalar values of the user rating matrix are fit by dot products of user feature vectors and item feature vectors which consumes much less space than the matrix itself. By formulating the matrix completion problem as a factorization technique in this way, unknown rating values of the matrix can be predicted in an accurate and efficient way.

Matrix factorization has a lot of variants including SVDFeature [1], SVD++ [2], MatRec [3], Zipf Matrix Factorization [4] and Alternating Least Squares [5]. The formulation of these problems is modification to the classic one and the solving techniques are mostly popular techniques in the academia and industry such as Stochastic Gradient Descent. Matrix factorization techniques are mostly shallow models. However in recent years, deep learning approaches such as Deep Matrix Factorization [6] have also been invented in the field, which has good accuracy performance, with much slower speed.

In some application scenarios, recommender system is a multi-task learning problem that not only optimizes the technical accuracy metric, but also other metrics such as serendipity and diversity. In the field of context-aware recommender system, people need to take consideration of other factors in recommending data. In these problems, classic matrix factorization and its popular variants are not suitable without major revision to the problem formulation.

The problem arises due to the simplicity structure of the problem formulation of matrix factorization. A simple scalar value fitted by dot products has very limited information that cannot incorporates information other than rating values such as the location, the weather, user's mood, etc. Some researchers resort to tensor factorization [7][8] to solve the problem, but its computational cost makes it implausible in real world applications. Therefore, we propose to replace the scalar values of the user rating matrix by matrices whose diagonal values are rating values and values in other places are side or contextual information.

In this paper, we propose a special case of our framework, i.e., replacing the rating scalars by matrices incorporating popularity information and fitting the matrices by matrix products of user feature matrices and item feature matrices. We illustrate the competence of our algorithm in accuracy metric and fairness metric in the experimental section.

## II. Related Work

Matrix factorizatoin is capable of excelling in rating prediction in recommender systems and its variants are popular in data science competitions and internet products. Most people modify the approach of the classic matrix factorization to improve the accuracy performance. Methods in this line include SVDFeature [1] and Alternating Least Squares [5]. Other researchers focus on the fairness problem and as a side effect find out that there is no trade-off between fairness and accuracy.

Tensor factorization [7][8][9] is a generalized version of matrix factorization. There have been some literature on its application in context-aware recommender system. It would be natural for people to speculate that tensor factorization is an ideal substitute for matrix factorization when side information is needed or multiple optimization goals are required. However, its space complexity is implausible in commercial environments, as we discuss in the Discussion Section.

Fairness is a serious issue of AI algorithms and has raised awareness in the academia in recent couple of years

[10][11][12]. Different researchers have proposed different evaluation metrics to measure the degree of fairness. In this paper, we borrow the idea called Degree of Matthew Effect from H.Wang's 2021 work [4] as the fairness evaluation metric for our framework.

### III. PROBLEM FORMULATION

Matrix factorization problem is essentially a scalar fitting problem. Due to the simplicity of the scalar values, it is very difficult to incorporate side information into the matrix factorization problem formulation without major revisions. But side information is ubiquitous in our daily lives. For example, industrial practitioners have been trying to develop mood-based or context-based music recommendation where side information such as user's mood or location is a must-have consideration factor in the model.

The classic matrix factorization is formulated as follows:

$$RMSE = \sum_{i=1}^{m}\sum_{j=1}^{n}(u_i \bullet v_j - R_{ij})^2$$

The RMSE is essentially the sum of squared error between the scalar value of the user rating matrix and the fitting dot product value.

To enhance the feasibility of matrix factorization, we replace the scalar rating values in the user rating value matrix as follows:

$$R = \begin{bmatrix} \begin{bmatrix} r_{1,1} & \cdots & \cdots \\ \cdots & \cdots & \cdots \\ \cdots & \cdots & r_{1,1} \end{bmatrix} & \cdots & \begin{bmatrix} r_{1,n} & \cdots & \cdots \\ \cdots & \cdots & \cdots \\ \cdots & \cdots & r_{1,n} \end{bmatrix} \\ \cdots & \cdots & \cdots \\ \begin{bmatrix} r_{n,1} & \cdots & \cdots \\ \cdots & \cdots & \cdots \\ \cdots & \cdots & r_{n,1} \end{bmatrix} & \cdots & \begin{bmatrix} r_{n,n} & \cdots & \cdots \\ \cdots & \cdots & \cdots \\ \cdots & \cdots & r_{n,n} \end{bmatrix} \end{bmatrix}$$

, where R is an n by m matrix ( n is the number of users and m is the number of items) and each element of R is a t by t square matrix whose diagonal elements are the scalar rating value and other positions are side or contextual information. In this way, the classic matrix factorization technique is reformualted as follows :

$$RMSE = \sum_{i=1}^{m}\sum_{j=1}^{n}\|U_i \bullet V_j - R_{i,j}\|^2$$

, where U, V and R are all matrices. We name this formulation MatMat, meaning matrix factorization by matrix fitting.

We investigate a special case of MatMat : We incorporate the popularity information of users and items into the target matrix in the hope to enhance accuracy and fairness since it has been shown lately that the popularity factors are crucial in the fairness problem of recommender systems. We define the matrix to be fitted below:

$$R_{i,j} = \begin{bmatrix} \dfrac{r_{i,j}}{\max(r)} & \dfrac{item\_rank_j}{\max(item\_rank)} \\ \dfrac{user\_rank_i}{\max(user\_rank)} & \dfrac{r_{i,j}}{\max(r)} \end{bmatrix}$$

, then U and V becomes n by 2 and 2 by m matrices. Notice there are 2 rating values in the matrix to be fitted, so in the end there will be 2 approximations of the scalar rating value. We take the average of the 2 approximations as the prediction of the scalar rating value.

To solve for the optimal matrices U and V, we adopt the popular optimization technique Stochastic Gradient Descent. Other techniques such as Adam are also applicable in the problem context.

As noted by researchers such as H.Wang [13], the popularity factor transfers or enlarges the popularity bias effect in the input data structures to the computational stage and the output data structures. The effect can be analyzed quantitatively in certain algorithms such as collaborative filtering. Matrix factorization techniques such as Lambert Matrix Factorization [14], MatRec [3] and Zipf Matrix Factorization [4] have been invented to tackle the fairness problem in a much more scientific way than years before.

In our MatMat formulation that incorporates popularity effect, we provide a more complex relation between popularity information and predicted rating scores. It's not linear as in some of the previously invented techniques. More interactions are involved between popularity information and rating scores. We hope in this way, we can obtain better prediction scores in both accuracy and fairness.

In the following section, we prove the competence of our algorithm in comparison with the classic matrix factorization and its variants. We extract a well-tested recommender system dataset and compare our MatMat formulation with the classic matrix factorization method.

### IV. EXPERIMENT

We use the small dataset of MovieLens that include 610 users and 9724 items to test our MatMat formulation with popularity information. Due to the randomness of Stochastic Gradient Descent, the results are different in different experiments. We select 2 of the experimental results and illustrate the data in figures. The metrics are Mean Absolute Error (MAE) for accuracy performance and Degree of Matthew Effect for fairness evaluation.

We do a grid search on the gradient learning step and illustrate the variation of performance according to the different parameters (Fig. 1, Fig. 2, Fig. 3 and Fig. 4).

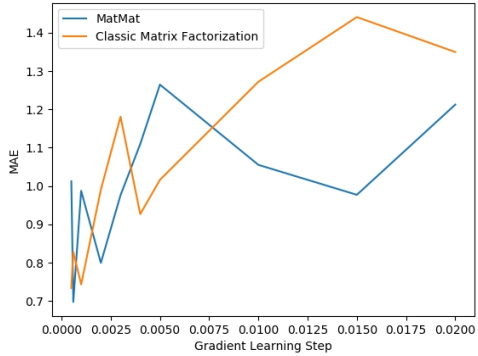

Fig.1 MAE comparison between Classic Matrix Factorization and MatMat in Experiment 0

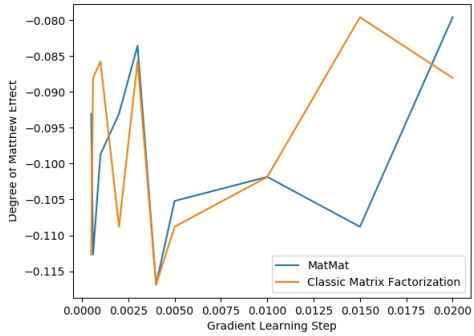

Fig.2 Degree of Matthew Effect comparison between Classic Matrix Factorization and MatMat in Experiment 0

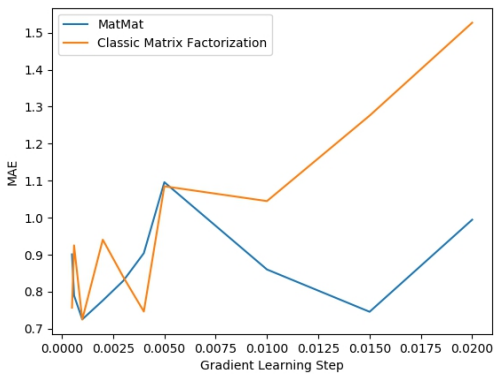

Fig.3 MAE comparison between Classic Matrix Factorization and MatMat in Experiment

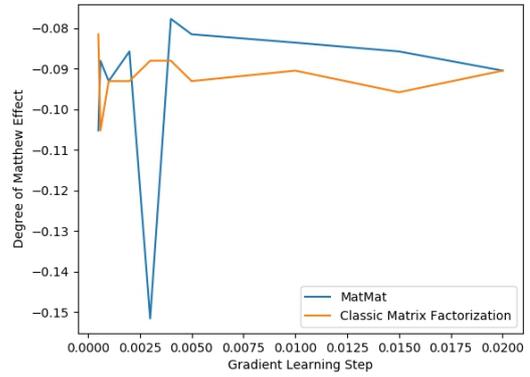

Fig.4 Degree of Matthew Effect comparison between Classic Matrix Factorization and MatMat in Experiment 1

The experimental results point to the conclusion that MatMat produces better results in both accuracy and fairness metrics in most parameter spectrum, with only very few exceptions. The method is more robust and precise and fair compared with the classic matrix factorization approach solved by Stochastic Gradient Descent.

## V. DISCUSSION

The formulation of MatMat framework is similar to Tensor Factorization. However, the theory is different and the computational complexity is also strikingly different on scales. For example, if we model our special case of MatMat framework with only item popularity information would cost more than 4.8 TB data storage to just store the input tensor. A recommender system for a website with hundreds of millions of users and tens of thousands of items would eliminate the possibility of its application in any company on this planet.

The initial idea of our MatMat framework was indeed inspired by tensor factorization, and we did try to solve the recommender system problem using tensor factorization. But since we only have commercial laptops and cloud services at hand, it was impossible to implement a full-fledged tensor factorization model with both the user popularity and item popularity (The data storage for the input tensor would be on the scale of Petabyte).

Tensor factorization is still not applicable in big data settings. Our method is a great substitute for tensor factorization with competitive performance and much lower computational cost in space (and possibly in time). MatMat is widely applicable in context-aware matrix factorization and multi-task learning.

## VI. Conclusion

In this paper, we propose a new framework for matrix factorization called MatMat. The framework replaces the scalar values of the user rating matrix by matrices incorporating side information such as popularity effect, location, user's mood etc. In this way, MatMat is capable of modeling more complex interaction among data other than linearity. Therefore MatMat has great potential in multi-task learning and context-aware recommender systems.

We prove in experiments the validity and competence of our techniques. In future work, we would like to explore the application of the MatMat framework in the field of context-aware recommender system and image processing.